# Are atomic clocks slowing down?     Possible connection between the quantum and the cosmos: A feasible experimental test of objective realism.

**R.**Ramanathan

Department of Physics,University of Delhi,Delhi,India.

Abstract: An alternative to the current accepted model of expanding universe is presented. The proposal is anchored in the objective realist, stochastic formulation of quantum mechanics wherein the Planck constant plays the role of a diffusion constant which need not be an universal constant for all epochs. This underlying noise is charecterised by a very high initial value of the Planck constant which gradually decreased over the epochs gone by, finally reaching its present value .It is this decrease in the strength of the Planck constant that causes a spectral red shift with the passage of time .The compatibility of this proposal with some cosmological data is shown.Feasible spectroscopic measurement of a specific spectral line frequency done over a period of a few years can decisively either validate or invalidate the proposal in a terrestrially bound experiment. This proposal predicts a slowing down of atomic clocks by nearly 2. $10^{-18}$ s with respect to every preceding second. A positive experimental outcome will validate the objective realist basis of quantum physics as advocated by Einstein and many others.

In a recent paper C.Wetterich [1] has proposed an alternative to big bang cosmology, which has created a lot of interest in the model.It augers well for the wellbeing of science in general and cosmology in particular. Unless such alternatives are discussed and assessed for their scientific robustness, there is a danger of going into a millennial intellectual slumber like the one science had in the aftermath of Ptolemaic geocentric model. After all, as Einstein famously said "physical principles are free inventions of the human mind", the essence of physics lies in testing the principles on a continuing basis, and this paper proposes just one more test.

Many years ago ,I had forwarded an idea[2] for an alternative to big bang cosmology, anchored in a possibility that arises in the stochastic objective realist formulation of quantum mechanics[3],which has the same consequences as Wetterich's  model, though by a totally different means.  Now that there is some interest in alternative ideas ,I present the proposal in a slightly altered form .One reason for rejecting such a proposal is that ,at some level, any variation of universal constants will imply a violation of local Lorentz invariance ,but such violation may well occur at some level.

The main observation of all objective realist stochastic formulations[3,4] of quantum mechanics is that the equations governing quantum processes and the probability interpretation for the square modulus of the quantum state can be viewed as a peculiar kind of diffusion process. At the simplest level, the nonrelativistic Schrodinger equation of a free particle of mass 'm' is,

$$\frac{\partial}{\partial t}\Psi = i\, h/2m\, (\partial/\partial x)^2\, \Psi \qquad (1)$$

It can be shown [3,4], that the above equation is rigorously derivable from certain generalized stochastic principles. The coefficient h/2m can then be identified as a diffusion constant. In this view of the quantum phenomenon, the diffusion process is triggered by some strange, non-local, gravitation like noise(noise with gravitational characteristics), whose strength is given by the value of the Planck constant. Assuming this noise is of cosmic origin, there is no compelling reason to assume that the value of the Planck constant has remained constant through all cosmic epochs. So, let us assume that the Planck constant has varied over cosmological times. This is not the first time that a variation of a universal constant is postulated, Dirac had speculated on the time variation of the Gravitational constant[5], and the time dependence of Planck constant was considered by many authors previously[6].These ideas have been noticed and discussed in the literature, but no further progress could be made as they have not thrown up any testable experimental signatures. On the other hand the present proposal makes an experimentally verifiable prediction.

## 2. Variable Planck constant as the cause of galactic red shift:

The big bang model rests on the interpretation of galactic spectral redshift as due to Doppler shift caused by galactic recession velocity. Further reinterpretation was made by Hubble by using the red shift measurements to arrive at a position velocity relation,

$$V = HR \tag{2}$$

Where v is the galactic recession velocity, H is the Hubble constant and R is the galactic separation. H is not a constant in time but as a first approximation we may assume its constancy.

Spectral lines generated by electronic transitions in an atom is governed by the Rydberg relation,

$$\nu = m\, e^4/2h^3 (1/n_1^2 - 1/n_2^2) \tag{3}$$

where ν is the frequency of the spectral line generated by the transition between levels $n_1$ and $n_2$.

When we observe distant galaxies we look into the past. The light from a galaxy distant R from us emitted it τ years ago,

$$R = c\tau. \tag{4}$$

In our model the Planck constant h(τ) was the value at the time the transition took place in the distant galaxy, so (3) is modified to,

$$\nu' = m\, e^4/2h(\tau)^3 (1/n_1^2 - 1/n_2^2) \tag{5}$$

From (3) and (5) we can derive the red shift formula

$$(\nu - \nu')/\nu = H\tau = 1 - h^3/h(\tau)^3 = z \tag{6}$$

which yields the relation,

$$h(\tau) = h/(1- H\tau)^{1/3} \qquad (7)$$

### 3. Consequences:

Equation (7) gives the change in Planck constant over cosmological times in the simplest model. This formula is independent of dimensions, and represents a scale independent change.

Therefore, at $\tau =1/H$, the so called age of the universe which at present is estimated to be 15 billion years, the planck constant would have an enormous value. It would have been a singular event at the beginning of the cycle of the universe, but there is no space time singularity that bedevils the big bang cosmology. Like in Wetterich's model the atoms are also getting smaller with the passage of time. The fine structure constant is also becoming larger as the universe ages. Equation (7) implies that a particular spectral line arising from an ancient atomic transition has energy $h(\tau) \nu'$, while its present day counterpart has the same energy given by $h\nu$, thereby making $\nu > \nu'$.

The cosmic microwave background radiation would also have been generated by the massive noise explosion that seems to have happened in this scenario.

For the future (7) must be replaced by,

$$h(\tau) = h/(1+ H\tau)^{1/3}. \qquad (8)$$

The above relation indicates that the Planck constant will reach half its present value a hundred billion years hence. As the universe ages quantum processes cease to exist and so will atoms, and elements, and the universe will revert back to its cold process less neutral phase till the great bomb maker creates a gravitation like noise yet again.

## 4.Compatibility with cosmological data:

**Though** this model shares certain features with the 'Tired photon' model of cosmology[7],it is very different from it because the mechanism for the red shift in this model is well defined, the latter model is quite ambiguous regarding the cause of tiredness of the old photons.One of the main refutation of the 'Tired photon' models is the incompatibility of the model to account for the lengthening of lifetimes of older supernovas with higher values of red shift ' z ' as compared with younger ones.In the big bang model the broadening of supernova decay time of older supernovas is attributed to time dilatation of photons emitted relative to the younger or nearer ones,and the broadening's' is proportional to (1+z) for z<1,so there is a linear rise 's' with increase in 'z' as indicated by data. On the other hand the 'Tired photon' model which leads to a non expanding universe predicts a constant value of 's' for all 'z' ,contrary to the data behavior.

In the present model the super nova decay is the result of collective resonant decay of matter present in it ,decaying with a mean decay time of 'h/Γ 'where  ' Γ ' is the energy width at half maximum of the decay rate curve. So, in this model the supernova decay time is proportional to h(τ) or 's' varies as $1/(1-z)^{1/3}$ ,the model therefore predicts a rise in 's' compatible with the data as shown by fig(1).

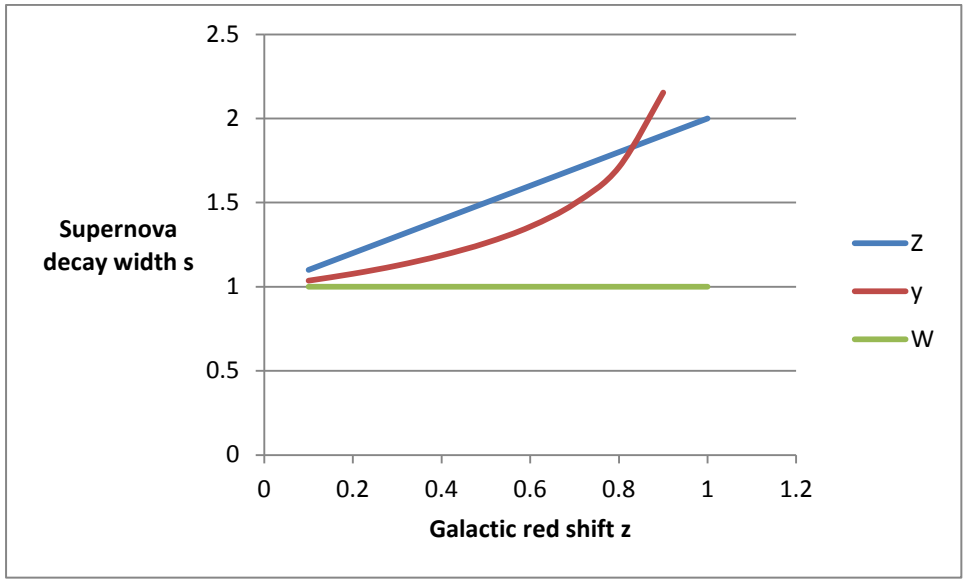

Fig(1): the comparative predictions of big bang model(blue line),tired photon model(green line) and the present model(red line).

## 6.Experimental test of the proposal:

In this proposal ,the galactic red shift is attributed to the declining strength of the Planck 'constant' with passage of time, with every specific atomic spectral line frequency suffering a red shift of the order of $10^{-10}$ per year,or of the order $10^{-18}$ per second,an order of magnitude less than the limiting accuracy of present day atomic clocks. Over a decade it will be of the order of $10^{-9}$ ,a red shift of this magnitude should be measurable using advanced spectroscopic techniques available today by keeping a specific laboratory atomic sample under continuous spectroscopic observation ,making continuous ,very high accuracy measurement of the frequency of a specific spectral line over the period of several years. A detection of red shift of this magnitude will validate the model, and also uphold the sanctity of 'objective realism' as a corner stone of science. A negative result, however, will only negate the present proposal.